# Suitability of using technical indicators as potential strategies within intelligent trading systems

Evan Hurwitz
University of Johannesburg
Faculty of Engineering
Doctoral Candidate
Johannesburg, South Africa
hurwitze@gmail.com

Tshilidzi Marwala
University of Johannesburg
Faculty of Engineering
Dean of Engineering
Johannesburg, South Africa
tmarwala@uj.ac.za

*Abstract*—The potential of machine learning to automate and control nonlinear, complex systems is well established. These same techniques have always presented potential for use in the investment arena, specifically for the managing of equity portfolios. In this paper, the opportunity for such exploitation is investigated through analysis of potential simple trading strategies that can then be meshed together for the machine learning system to switch between. It is the eligibility of these strategies that is being investigated in this paper, rather than application. In order to accomplish this, the underlying assumptions of each trading system are explored, and data is created in order to evaluate the efficacy of these systems when trading on data with the underlying patterns that they expect. The strategies are tested against a buy-and-hold strategy to determine if the act of trading has actually produced any worthwhile results, or are simply facets of the underlying prices. These results are then used to produce targeted returns based upon either a desired return or a desired risk, as both are required within the portfolio-management industry. Results show a very viable opportunity for exploitation within the aforementioned industry, with the Strategies performing well within their narrow assumptions, and the intelligent system combining them to perform without assumptions.

*Share; Portfolio; Risk; Technical Analysis; Temporal Difference; Energy function; Data generation; agent*

## I. INTRODUCTION

In order to investigate the potential for applying machine learning to the tasks of equity trading and portfolio management, one must first identify precisely what new strengths the tool brings to the party. In the case of machine learning it is the abilities of the system to adapt to changing circumstances while exploiting learned knowledge [1], and to learn untutored and online [1], hence in many ways becoming free of investigator bias [2]. Since this choice is the strength of the system, strategies need to be developed for the machine learning system to choose between. For these, popular strategies have been taken from the field of Technical Analysis, since the strategies have well-defined assumptions that can be used to test the ability of the machine learning system to appropriately adapt to given underlying patterns within the data. In order to generate the data, appropriate methods of generating noise are investigated, since actual equity prices fluctuate with an appreciable amount of noise [3]. The trading of these generated shares (pseudoshares) is evaluated according to their applicability within the machine learning environment, and not towards the needs of the portfolio management industry, which contrary to popular belief is not in fact to obtain the highest possible return, but rather to maximise return for a given value of *risk* [4], which would be the ultimate goal of applying the results determined within this paper. This research has been undertaken as a portion of the research in pursuit of a Doctoral thesis at the University of Johannesburg.

## II. TECHNICAL ANALYSIS

Technical Analysis has received a substantial amount of criticism over the years [5]. This criticism has been, for the most part, entirely justified. The primary criticism of technical analysis is that all of the methods make broad assumptions about the underlying share price (and patterns therein) without perform any verification of those assumptions [6]. This criticism becomes especially damning when noting that many methods not only make these assumptions, but in fact make contradictory assumptions – assumptions that are in fact mutually exclusive. Fortunately, it is precisely these facets of technical analysis that makes techniques therein appropriate for the purposes of this investigation. By choosing appropriate methods and creating specific functions it is possible to test the appropriateness of these methods on data both with and without the underlying required characteristics, and once the success of appropriate methods is established to then test the ability of a machine learning agent to successfully choose between measures for appropriate exploitation of a given, unknown signal. The two methods chosen to be investigated are those of the Moving Average Crossover-Divergence (MACD) indicators and the Relative Strength Indicators (RSI).

### A. MACD

In order to construct a MACD indicator, one first needs to become familiar with the concept of *moving averages* [6] [7]. Moving averages are calculated in two ways, namely simple moving averages and exponential moving averages [6] [7]. A simple moving average is constructed according to the following formula [7]:

$$MA(\tau) = \Sigma P_\tau / \tau \qquad (1)$$

Where τ is the number of days of moving average to use, and P is the vector of prices leading up to the current price. A cursory examination of (1) indicates that it in fact approximates a form of a low-pass filter, with varying degrees on smoothing (and consequent lag in responsiveness) depending on the variable τ. Weighting factors can be placed on the formula in order to quicken responsiveness to changes, although the basic premise remains the same [7]. A common alternative is the exponential form, which is incrementally calculated as follows [8]:

$$EMA_t = EMA_{t-1} + \alpha(P_t - EMA_{t-1}) \qquad (2)$$

The parameter α in this case is a weighting factor determined by the number of day's memory to effectively keep, as follows [8]:

$$\alpha = 2/(N + 1) \qquad (3)$$

This equation will be more familiar as an infinite impulse response filter, with the added advantage over (1) being that all information is in fact retained, although data points' importance is gradually bled away as it becomes less recent.

The idea behind the MACD indicators is to use two moving averages, one faster (i.e. say a 9-day moving average) and one slower (so for example a 50-day moving average). When one crosses the other, a signal is created to buy or sell depending on whether the faster signal is greater than or less than the slower signal [9].

The underlying assumption here is that the share itself is following a "trend" – that is, some approximation of a straight-line curve, and hence any change in trend will be indicated by the crossing over of the slower and faster filters.

### B. RSI

The RSI indicator is designed to take advantage of cyclical data [10]. The equation for the RSI indicator is as follows [10]:

$$RSI = 100 - (100/1 + RS) \qquad (4)$$

Where RS is the average of X days price rises / X days price drops. This is used in conjunction with setting specific buy and sell points, typically 30 and 70 respectively, in order to take advantage of the cyclical nature of the share [10].

The implicit assumption in this method is that the share has a repeating cycle, and that when the price has risen it is then set to drop, and vice-a-versa.

### III. DATA GENERATION

In order to generate usable pseudodata, the following facets need to be addressed in the generation thereof:

- An underlying pattern needs to be able to be placed within the generated data.
- Noise needs to be added to the data to simulate market conditions.

The pseudodata generated were done according to the following specifications.

### A. Patterns

In order to evaluate the efficacy of the trading strategies presented, data needs to be created that fits each respective strategy. In order to fit the MACD approach, data sets with discernable trends are created, while data sets with definite cycles are created in order to utilize the RSI-trading strategy.

*1) Trending data*

In order to create trending data, data is created according the well-known straight-line equation:

$$y = mx + c \qquad (5)$$

By modifying the equation parameters at random points in time, the signal can be kept continuous while changing the trend within the function. When modifying the parameter m in (5) to change the trend, the parameter c must be recalculated in order to maintain continuity. (A quick note on continuity – while it is true that the values are in fact discrete price values, the discrete value at the point of transition should still be consistent, and hence the term 'continuous' in this context). The results of such a generating scheme can be seen in Figure 1

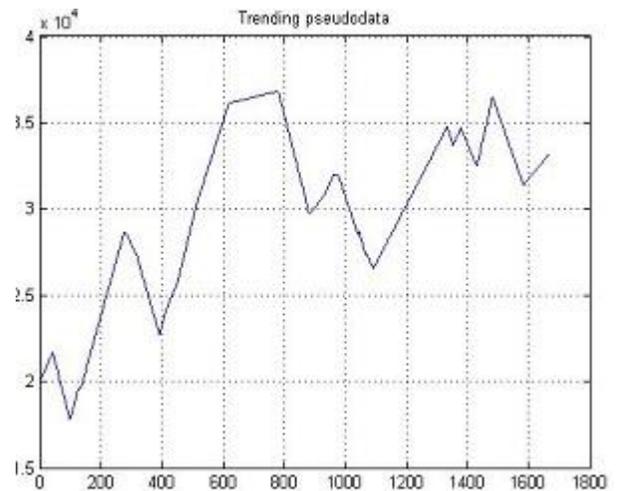

Figure 1. Trending Pseudodata

*2) Cyclical Data*

In order to create cyclical data, data is created using a simple sinusoid:

$$y = A(\sin(bx)) + c \qquad (6)$$

As with the trending data, changes in time are introduced by changing parameters (in this case A and b) while calculating c to once again retain continuity. The results of such manipulations can be seen below in Figure 2, where long periods have been used.

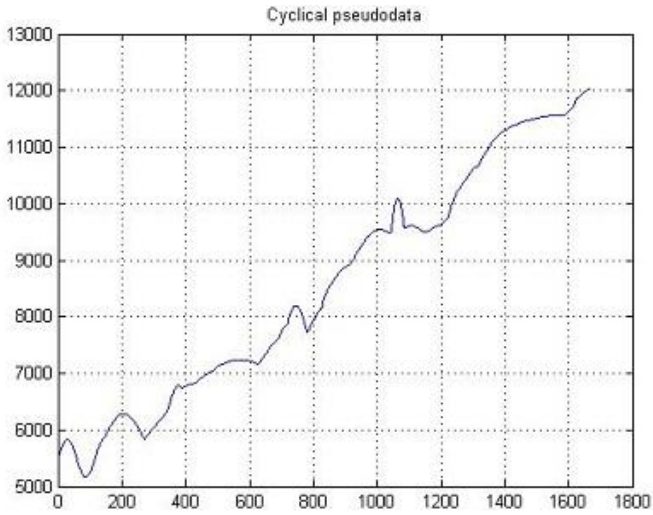

Figure 2. Cyclical pseudodata

## B. Noise

While many methods exist to introduce random noise into an equation, two were investigated in this research. The first, and most obvious, is to simply overlay a random signal over the base signal. The second is to create a noise signal based on random fluctuations, but to influence those fluctuations with an energy signal, giving the noise more freedom while still keeping it constrained to the underlying signal.

### 1) Simple random noise

The simplest method of producing noise is to produce a random value R from zero to one at each point, and apply the following formula [11]:

$$Xi = \alpha Xi (1 + Ri) \qquad (7)$$

In equation 7 $\alpha$ is the maximum percentage noise allowable in your given signal. The problem with this method is that it produces overly randomised data. The problem with such data is that actual share price movements form facets that can be construed as underlying patterns (or may even be the result of smaller patterns) and no such facets emerge using this form of noise generation.

### 2) Energy-signal noise

In order to produce noise that is more realistic, this research investigated the use of an energy function in the generation of noise for an existing signal. The concept is to allow the signal to be fed the previous value, and make a random movement for its proceeding value. The difference between that value and the 'clean' value at that point will then influence the random movement of the next stage, retaining the underlying pattern. The governing equations look as follows:

$$X_{t+1} = X_t + \alpha E(Xt) \qquad (8)$$

$$E(Xt) = R + K(X_{t-1}/Y_{t-1}) \qquad (9)$$

Where K is a constant determining the maximum allowable noise, and Y is the original, 'clean' signal.

This form of noise proved to produce facets of a similar nature to those found in actually data, as illustrated in Figure 3, which depicts the original function and the 20-day moving average (hence filtered) noisy function produced with the energy noise function, and as such was deemed a better tool for producing noise than the simple random noise generation discussed in II.B.1.

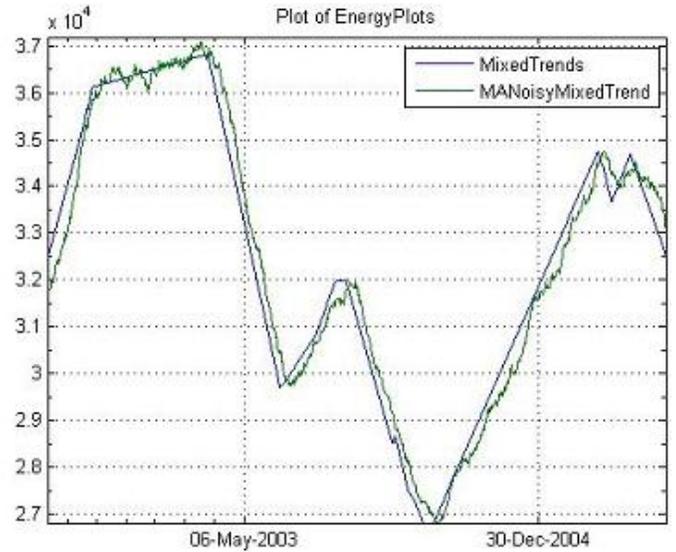

Figure 3. Original function compared to noisy function

## IV. EXPLOITATION VERIFICATION

In order to test the efficacy of the trading strategies, Data was created as described above for each trading strategy. The data was then back tested against the generated pseudodata, with noise and without, both against the data that contains the underlying assumptions made by the trading strategies, and against data that does not contain the underlying assumed patterns, in order to verify the success and / or failure of the strategies in exploiting the facets of their signals. These results are all compared to a buy-and-hold strategy in order to determine if the actual trading itself had any real payoff.

### A. MACD Results

The initial results, looking at clean data, are very promising. These are illustrated in figure 4, showing the base price and the generated indicators used by the trading system to trade on the share price. Comparing to the Buy-and-Hold strategy illustrated in figure 5, we see that the trading system outperforms the buy-and-hold buy successfully avoiding downturns in the share price. (It is worth noting that while the strategy makes no gains during this time, it would make gains if it had the ability to 'short' shares, which for simplicity's sake has not been implemented)

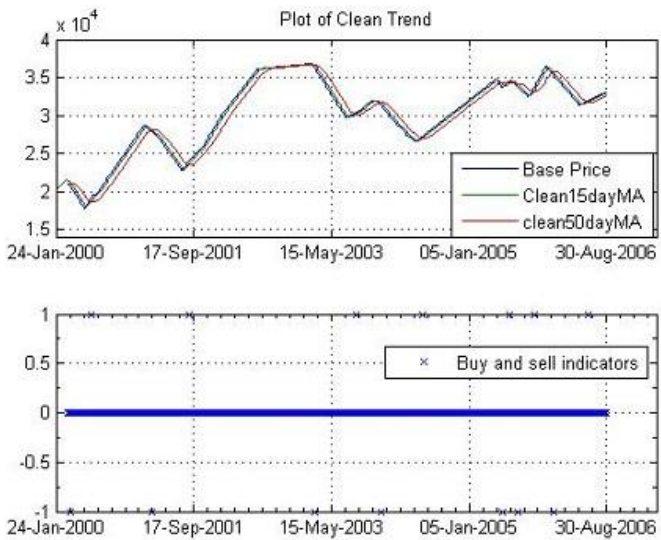

Figure 4. MACD trading on clean trending data.

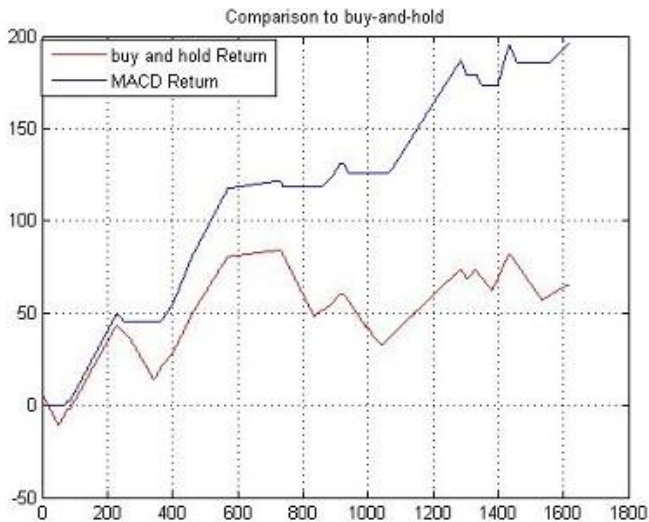

Figure 5. Buy-andHold comparison for clean MACD

While these results look promising, the results once noise is added to the signal do change things somewhat. While still performing substantially better than the Buy and Hold strategy, the MACD return drops from 195% return over the time period to 155%. This drop is purely due to the noise added, noise of only 10%. This result shows a susceptibility to being over-sensitive to high-frequency noise, something that can in fact be catered to by adjusting the moving-average parameters to create what are effectively more fine-tuned filters, but of course the danger is in tuning out the higher-frequency components of the underlying function, which would slow the response of the strategy and adversely affect its performance as well.

Applying the same trading strategy to the oscillating function shown in figure 6 returned erratic results that were unsuccessful in identifying opportunities for exploitation, and also badly under performed when compared to the buy-and-hold strategy. This result is not unexpected, and indeed fits the hypothesis that the MACD indicator is appropriate for signals with underlying trending functions while being inapplicable to signals without the underlying trends.

### B. RSI

A similar process was undertaken to evaluate the efficacy of the RSI strategy. The same basic function shown in figure 6 was used for the strategy to attempt to exploit.

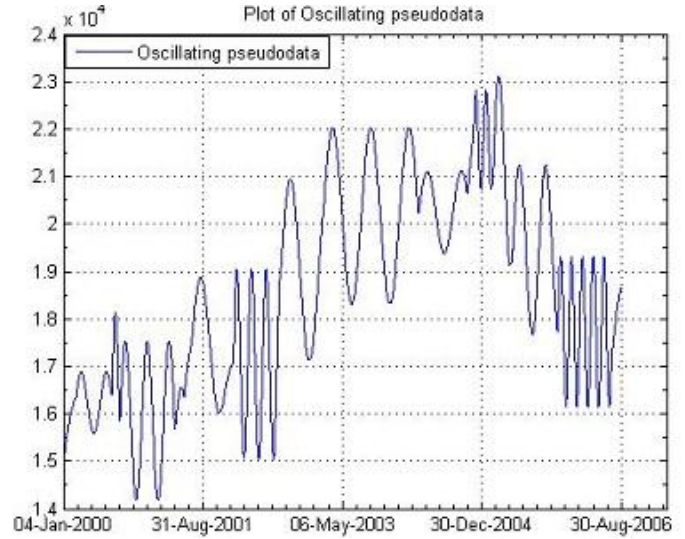

Figure 6. Clean oscillating pseudodata

The results of this trading experiment were far more erratic than seen in the MACD trading strategy. The defining variable of the RSI indicator, namely the day's history taken into account, determines the frequency to which the strategy is sensitive. As such, the trading strategy performed very erratically on the mixed oscillator pictured in figure 6, owing to the varying frequencies for which a static strategy could not account. (It is worth noting that changing gradients did not hamper the MACD strategy in the same manner, showing it to be a more robust method of exploiting its particular assumption). Still, the RSI did manage to outperform the buy-and-hold strategy.

Performance on noisy data was far less promising. While a positive result was obtainable, it required extensive manipulation of the trading variable, so much that it can be considered more of a local minimum than a generic trading tool. The strategy performed acceptably with noise provided it was given a set of pseudodata with only one frequency, which it could handle without problem, although also with a loss of return owing to noise interfering with the indicators, as was observed with the MACD strategy.

Performance on the trending pseudodata returned poor results, once again as expected. Since the trending pseudodata had no harmonics built in, the lack of performance thereupon is not surprising, and reinforces the earlier hypothesis.

## V. Conclusions and recommendations

For the most part the experiment was a success. The generation of noise provided adequate obfuscation to the strategies to not be considered trivial, while still retaining the underlying patterns that the respective strategies assumed are present. In the clean cases, both strategies were able to take advantage of their underlying assumptions while unable to perform when those assumptions were removed. The MACD strategy was able to perform even under noisy conditions, showing it to be a robust strategy, very much appropriate for use within a larger machine learning system. The RSI indicator, on the other hand, proved to be very finicky when presented with reasonable challenges, and as it stands is not suitable for use within a larger intelligent agent. It shows promise with the recognition and exploitation of its underlying assumption when present, but in order to be included as part of an intelligent agent needs to incorporate some method of detecting appropriate variable values in order to become more robust, and hence usable within a machine learning agent.